\def\year{2022}\relax
\documentclass[letterpaper]{article} 
\usepackage{aaai22}  
\usepackage{times}  
\usepackage{helvet}  
\usepackage{courier}  
\usepackage[hyphens]{url}  
\usepackage{graphicx} 
\usepackage{natbib}  
\usepackage{caption} 
\DeclareCaptionStyle{ruled}{labelfont=normalfont,labelsep=colon,strut=off} 
\usepackage{algorithm}
\usepackage{algorithmic}
\usepackage{amsmath}

\usepackage{booktabs}

\usepackage{newfloat}
\usepackage{listings}
\floatstyle{ruled}
\newfloat{listing}{tb}{lst}{}
\floatname{listing}{Listing}
\title{SMI-5: Five Dimensions of Social Media Interaction for Platform (De)Centralization}
\author{
    Lynnette Hui Xian Ng\textsuperscript{1}, Samantha C. Phillips\textsuperscript{1}, Kathleen M. Carley\textsuperscript{1}
}
\affiliations{
    \textsuperscript{\rm 1}    Center for Informed Democracy \& Social - Cybersecurity, Carnegie Mellon University\\
    Pittsburgh, PA, United States
}

\begin{document}

\maketitle

\begin{abstract}
Web 3.0 focuses on the decentralization of the internet and creating a system of interconnected and independent computers for improved privacy and security. We extend the idea of the decentralization of the web to the social media space: whereby we ask: in the context of the social media space, what does ``decentralization" mean? Does decentralization of social media affect user interactions? We put forth the notion that decentralization in the social media does not solely take place on the physical network level, but can be compartmentalized across the entire social media stack. This paper puts forth SMI-5: the five dimensions of social media interaction for describing the (de)centralization of social platforms. We then illustrate a case study that the user interactions differ based on the slices of the SMI layer analyzed, highlighting the importance of understanding the (de)centralization of social media platforms from an a more encompassing perspective rather than only the physical network.
\end{abstract}

\section{Introduction}
The decentralized web infrastructure emerged as a viable solution to regain some control from the central powers of web application, providing users with democracy, censorship, bandwidth and security \cite{alabdulwahhab2018web}. Among the applications on the web are social media platforms.

Social media are digital interactive technologies for creating, sharing and discovering content and interests between virtual social media users. Decentralized social network structures are a growing trend in the social media landscape, with a focus on privacy and community \cite{la2022information,jeong2024user}. These include decentralization using techniques like hosting separate instances (i.e., Mstodon, Discord), blockchain-based communication (i.e., Steemit, Versaity) or a Peer2Peer communication method (i.e., Secure-Scuttlebutt, Aether).
Research studies on decentralized social media platforms largely focused on microblogging platform Mastodon. Constructed using open-source protocols where administrators host separate instances for their communities, there is a greater sense of spontaneous connectivity between users in the Mastodon network due to the absence of boosting recommendation strategies that centralized social media has \cite{jeong2023exploring}, and that users networks exhibit high modularity and high number of communities \cite{la2022information}. Analysis of the network structure of Mastodon and Twitter reveals that bidirectional relationships are more likely as compared to Twitters, and that the presence of spambots is marginal, thereby showing that advantages of a decentralized social media: to strengthen user ties and reduce spambots \cite{zignani2018follow}.

Existing literature offers only a partial view of (de)centralization architecture as understood by a network topology perspective. It portrays only the most apparent difference of decentralized platforms, that is, the network topology \cite{zignani2019footprints}. \citet{zulli2020rethinking} redefined the social media stack in terms of topology, abstraction and scale to define Mastodon's unique sociotechnical configuration. Yet, the social media ecosystem is more complex than that, spanning users, interactive actions, content, the physical and network topology of the web-based platform application; each of which has a form of (de)centralization setup.
In this paper, we propose the SMI-5 framework, consisting of 5 dimensions of social media interactions as foundations of a social media platform. This framework is a generic description of communication flow between social media platforms. Elements of each dimension can be classified either as decentralized or centralized, illustrating that the concept of decentralization in the social media realm involves a high degree of complexity. Unlike the OSI-7 layers for computer networking \cite{day1983osi} in which layers peel back like an onion, the dimensions in SMI-5 are interwoven together. A single social media platform can be formed with both centralized and decentralized dimensions, providing unique features to the platform.

\section{What is ``Decentralized Social Media"}
\begin{figure*}  
    \centering
\includegraphics[width=\linewidth]{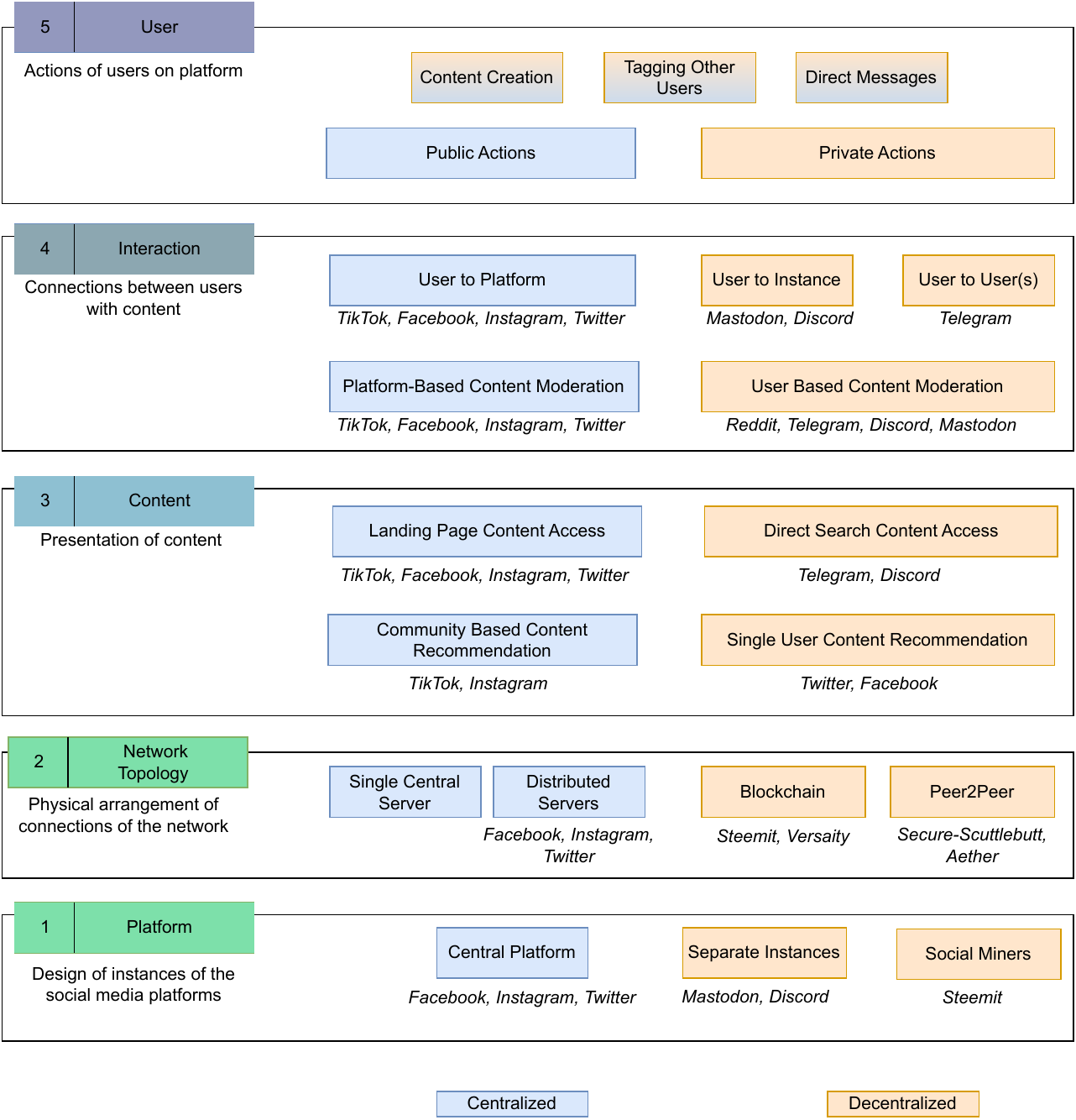}
    \caption{\textbf{SMI-5} Five Dimensions of Social Media Interactions for centralized \& decentralized layouts.}
    \label{fig:smi-layers}
\end{figure*}

The term ``decentralized social media" intuitively might refer to the network topology or the distribution of platform servers. Here, we put forth the notion that decentralization spans the entire social media interaction stack, from the instantiation of the platform, to the physical network layer, to the user actions. 
We abstract the social media stack into a frame of five separate layers. Therefore, the perspective of decentralization of social media relies largely on the slice of investigation. Figure \ref{fig:smi-layers} illustrates the 5 layers of Social Media Interactions (SMI-5). Below, we describe each layer of the SMI-5 stack.

\paragraph{Dimension 1: Platform} A centralized platform such as Facebook has a single service provider, managed exclusively by one entity, (i.e., Facebook is managed by Meta).
The key innovation of a decentralized platform in this dimension creates community-centered spaces, where people can setup independent instances, or servers, to build their local community \cite{10.1145/3355369.3355572}. Users in each instances cannot communicate directly with each other. Instances on platforms like Mastodon are setup with open-source code, and instances like Discord rely on an interconnected product. While decentralization in the platform dimension provides greater flexibility and control towards users and administrators of instances, it does have infrastructure-driven challenges with hosting separate instances \cite{10.1145/3355369.3355572}.

\paragraph{Dimension 2: Network Topology} This dimension refers to the physical arrangement of connections in a network. Many platforms make use of the distributed servers structure, where there are multiple servers, sometimes in different geolocation, working together to support the application. Distributed social media structure make use of Peer2Peer networks. One such example is Secure-Scuttlebutt which relies on partitioning the larger data to only the data that a user is interested in for each user \cite{tarr2019secure}. Blockchain networks is a type of Peer2Peer network that is built on trustless transactions \cite{alabdulwahhab2018web}. 

\paragraph{Dimension 3: Content} Centralization of content presents itself by the curation of content access and content recommendation. Platforms like TikTok, Facebook and Instagram presents their content through the landing page. This page is called the ``For You Page (FYP)" on TikTok, and the ``Feed" on Facebook. Decentralization of the landing page involves direct search content access, which requires users to deliberately search for their communities. Such decentralization can offer more expansive participatory cultures, for it provides a reflective mode of socio-technical organizing, where users purposefully decide their communities, and thereby, feed \cite{mannell2022alternative}. Centralization of content recommendation involves community-based recommendation systems, which is evident in TikTok, where the FYP is curated based on popular content around the user's region. Single User Content Recommendation systems make use of a user's past content interactions to promote similar content in the region.

\paragraph{Dimension 4: Interaction} This dimension forms the visible user connections. One aspect is the user-to-user connection. This includes a centralized user-to-platform structure (e.g., TikTok, Facebook), or decentralized structures like the user-to-instance structure (e.g., Mastodon, Discord), or a user-to-user structure (e.g., Telegram).
Another aspect is the user-to-content interaction, in particular in terms of content moderation. Centralized platform have their content moderated by the platform itself, which decides which posts and users are malicious usually with a black-box algorithm. 
Some work suggests that decentralized social media increases user privacy and reduces censorship if content moderation is performed at the user level \cite{paul2014survey}.  

\paragraph{Dimension 5: User} This dimension is where the user interacts and performs actions with the social media platform. A user can be a human user, or an automated bot user. Most social media platforms offer the same sets of actions, broadly in the categories of content creation, tagging other users or posting direct messages. These actions can be both public (centralized) or private (decentralized), regardless of platform.

\section{Differing Views of Interactions}
The structure of social media influences user relationships, and thereby influences our discovery of users. Decentralized platforms restructures social interactions through their unique layers. We illustrate the differences between analyzing user connections at the Interaction dimension, comparing the centralized platform Twitter where users post to the platform, and the decentralized platform Telegram, where users post to other users in a channel or a group.

The interaction dimension impacts how analysis is performed on social media data. Much of social media data analysis is performed by constructing focused relational networks where data from the User dimension is represented, for this view is a presentation of the user-to-user relationship. Here we present a brief case study to illustrate how actions in the user dimension present differently based on the architecture of the interaction layer. We also compare the interactions between automated bots and humans, and show that the discovery and analysis of these actions differ based on the view of the social media architecture.

\subsection{Data}
We obtained datasets surrounding the 2019 coronavirus pandemic for this illustration. Telegram data was collected using snowball sampling from the channels of a seed set of prominent disinformation spreaders during the coronavirus pandemic \cite{ng2024exploratory}. Twitter data was collected based on conspiracy theories surrounding the coronavirus pandemic \cite{ng2021coronavirus}. We apply the BotBuster algorithm to differentiate bot/human users based on an agglomeration of random forests \cite{ng2023botbuster}. Twitter and Telegram expose different sets of user features, and this bot detection algorithm uses features specific to each platform, that are also trained on data collected and annotated from the platforms. For example, Twitter exposes global retweet counts of tweets, whereas the decentralization of Telegram at the Interaction Layer means that the sharing metric of a post is per channel, so the number of forwards for each post needs to be cumulatively added across channels.

\begin{figure*}  
    \centering
\includegraphics[width=\linewidth]{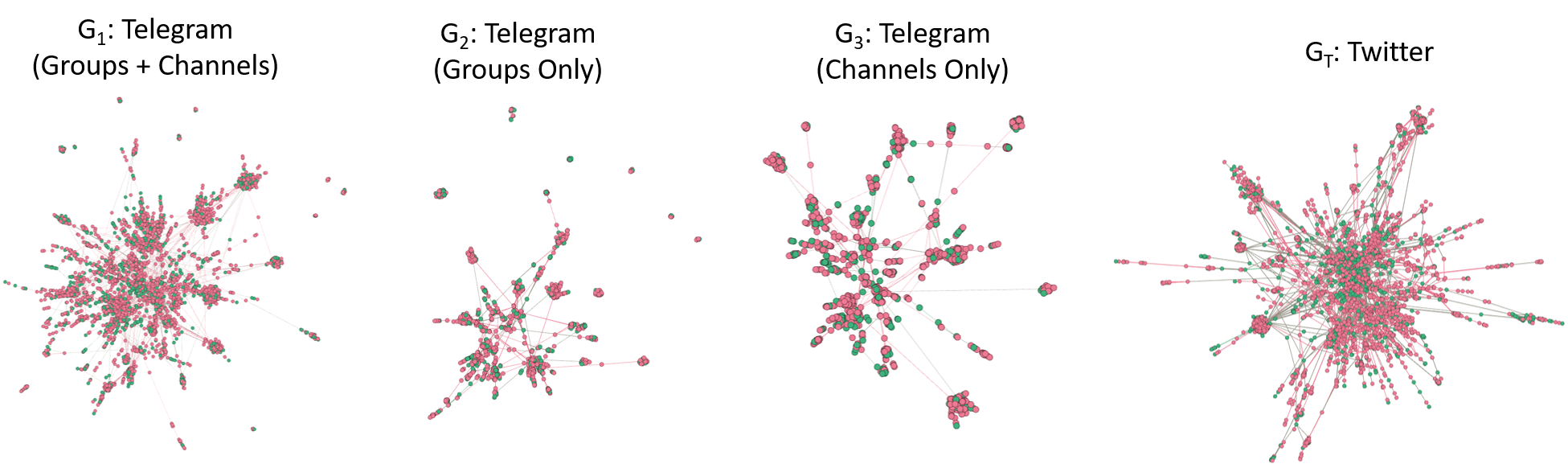}
    \caption{\textbf{Interaction Network Graphs.} Pink represent bot users, green represent human users. Links between two users represent an interaction (e.g., retweet, forward). A centralized platform like Twitter only has a single interaction view, while a decentralized platform has multiple views.}
    \label{fig:interactions}
\end{figure*}

\begin{table*}
    \centering
    \begin{tabular}{ccccc}
    \toprule
        \textbf{Network Metric} & \textbf{Twitter} & \textbf{Telegram (Channels)} & \textbf{Telegram (Groups)} & \textbf{Telegram (Groups + Channels)} \\     \midrule
        Num users & 3428 & 2111 & 2111 & 5408 \\ 
        \% bots & 41.56 & 27.00 & 27.00 & 27.29 \\ 
        Total Degree & 68.60$\pm$103.70 & 40.89$\pm$36.12 & 40.88$\pm$36.12 & 56.33$\pm$62.47 \\
        Betweenness Centrality & 1E-6$\pm$2E-6 & 1.24E-3$\pm$5.74E-3 & 1.34E-3$\pm$5.74E-3 & 6.48E-4$\pm$3E-3 \\ 
        Eigenvector Centrality & 3.99E-3$\pm$2.36E-2 & 2.61E-3$\pm$2.71E-2 & 2.61E-3$\pm$2.72E-3 & 1.11E-3$\pm$1.83E-2\\  
        Page Rank & 2.92E-4$\pm$1.91E-3 & 4.74E-4$\pm$1.54E-3 & 4.74E-4$\pm$1.54E-3 & 1.85E-4$\pm$7.44E-4\\ 
    \bottomrule
     \end{tabular}
    \caption{\textbf{Network Statistics of Interaction Networks.} The statistics are different depending on the point of view in a decentralized network, while there is only one view for a centralized network. }
    \label{tab:table_stats}
\end{table*}

\subsection{Constructing Interaction Networks}
Next we construct interaction networks for the Twitter and Telegram platforms. These two platforms differ mostly in their Interaction dimension: Twitter is centralized for public messages are posted in a user-to-platform fashion, while Telegram is decentralized for messages are posted in a User to User fashion either to user groups or to user channels. We construct network graphs $G=(V,E)$ of the user communications $\{A\}$ of these platforms. The set $\{V\}$ are user nodes and the set $\{E\}$ are edges where the users interacted with each other with a communication $A$. In Twitter, $A=\{\text{retweet}, \text{quote}, \text{mention}\}$. In Telegram, $A =\{\text{forward}, \text{reply}\}$. Further, Telegram has channels and groups, which users can converse within and in between. Therefore, Telegram presents three views: $G_1 = \text{Telegram (Groups + Channels)}$, $G_2 = \text{Telegram (Groups)}$ and $G_3 = \text{Telegram (Channels)}$. 
Twitter, on the other hand, has only a central platform which users reside. Therefore, there is only one main view, $G_T$.

We then zoom into the largest visual cluster and present them visually in Figure \ref{fig:interactions}. The view of $G_1$ is not merely a combination of the disparate views $G_2, G_3$, but has a form of its own. $G_2$ shows isolated communities, while $G_3$ are interconnected. Within these slices of $G_2, G_3$, we find slices of communities that are dominated by bot users, while $G_1$ and $G_T = \text{Twitter}$, bots are dispersed through the network. 

We further analyze the network statistics of the interaction networks are presented in Table \ref{tab:table_stats}. The proportion of bots on Twitter is $\sim$1.5x more than the Telegram views. The larger proportion of bots on the centralized Twitter platform could indicate that such a platform structure is easier for bots to operate on, and that the decentralization of Telegram aids in obscuring bots. While the proportion of bots on different views of Telegram interactions are similar, the other network centrality measures differ between the views. 

This case study illustrates the different user interaction structures that come from a decentralized dimension. Telegram, being decentralized in the Interaction dimension, shows that when considering the interactions with Channels, users have lesser influence on each other (i.e., lower eigenvector centrality values), than when considering the interactions in terms of Groups. This draws on the importance of giving thought to the analysis perspective, especially when examining user behavior along a decentralized dimension. Decentralization of a dimension results in multiple viewpoints, which thus can results in multiple analysis results.

\section{Discussion}
This paper puts forth the concept that decentralization in social media is not limited to the physical architecture and network paradigm but involves different perspectives. This work discusses five dimensions of decentralization that can be used in categorizing social media platforms, introducing the idea that decentralization may emerge at different granularity through different technologies. Dimensions of decentralization interact together to emerge new social media interaction effects that cannot be explained by considering only a single conceptualization of decentralization based on network architecture. For example, information diffusion on a platform with discrete communication channels (Dimension 4) will be faster and farther if the platform also has landing page content recommendations (Dimension 3).

Decentralized social media platforms provides hope towards community building and censorship concerns. Instead of having a central landing page where content recommendations from the platform's algorithms are pushed to the user, decentralized platforms rely on users to specifically search for their community, thereby providing a more curated user feed. Platforms such as Reddit, Discord and Mastodon mostly rely on user based content moderation \cite{bono2024exploration}, where community administrators safeguard the social norms of the group.

On the other hand, the decentralization of the social network poses social cybersecurity challenges in identifying bot interactions \cite{carley2020social}. The short study in this paper demonstrates the differences in discovering and differentiating the interactions in both centralized and decentralized social media. In particular, decentralization obscures interaction analysis. The social media analyst will require multiple planes of analysis for the decentralized network compared to a single plane in order to understand the social connections between users. Analysis of user behavior for discovery of users communities and information spread will require development of new techniques to perform analysis across singular and multiple planes, and aggregate them. Understanding the dimensions of social media interactions will enable development of new techniques to perform analysis across singular planes. While studying user behavior along decentralized dimensions, a key question should also be evaluated alongside: are the results from multiple analysis planes comparable? Can these multiple results be aggregated together?

As social media platforms evolve and the set of social media interactions expands, the SMI-5 Framework has to be updated. The makeup of these dimensions also needs to be iterated as newer creative forms of web interactions emerge, changing user interactions within social media platforms. 

Several lines of work can stem from this SMI-5 Framework: How are (de)centralized are each social media platform mapped across the dimensions? How can we accurately represent users, interactions and content across both dimensions, especially decentralized dimensions? How do dimensions interweave together to constitute a social media platform? How do cross-platform interactions take place between centralized and decentralized dimensions?



\section{Concluding Remarks}
This paper unpacks the complexity of the social media platform. It profiles interaction networks of different social media architecture, and will be extended with greater in-depth studies of user relationships between different centralization techniques within each SMI layer. This begins by breaking down the social media architecture into five interconnected dimensions, where within each layer (de)centralization can be defined differently. We then illustrate that the visualization of user layer interactions through the presentation on the interaction layer differs for different interaction architectures. Decentralization presents several view ports, while centralization presents a single view port, resulting in multiple facets. In terms of identifying automated, possibly malicious, bot users, the presentation of multiple facets means that multiple analysis layers are required, for these bots can leverage decentralization to obfuscate themselves, thus increasing the challenge of studying a decentralized social media network. We hope our discussion of layers of decentralization of a social media network will spur discussions towards defining a decentralized social media is.

\bibliography{aaai22}

\section{Acknowledgments}
This material is based upon work supported by the U.S. Army Research Office and the U.S. Army Futures Command under Contract No. W911NF-20-D-0002, Office of Naval Research (Bothunter, N000141812108), US Army Scalable Technologies for Social Cybersecurity (W911NF20D0002), Air Force Research Laboratory/CyberFit (FA86502126244), Office of Naval Research Scalable Tools for Social Media Assessment (N000142112229). The content of the information does not necessarily reflect the position or the policy of the government and no official endorsement should be inferred.

\end{document}